\documentclass{article}
\usepackage{ijcai20}

\usepackage{times}

\usepackage{soul}
\usepackage{url}
\usepackage[hidelinks]{hyperref}
\usepackage[utf8]{inputenc}
\usepackage[small]{caption}
\usepackage{graphicx}
\usepackage{amsmath}
\usepackage{booktabs}
\usepackage{color}
\usepackage{amsthm}
\usepackage{amsmath}
\usepackage{amssymb}
\theoremstyle{definition}
\newtheorem{definition}{Definition}[section]
\usepackage{amsthm}
\newtheorem{theorem}{Theorem}
\usepackage{mathtools}
\usepackage[T1]{fontenc}
\usepackage{comment}

\usepackage{latexsym} 

\title{Permissioned Blockchain Revisited: A Byzantine Game-Theoretical Perspective}

\author{
Dongfang Zhao\\
\affiliations
University of Nevada, Reno and University of California, Davis\\
\emails
dzhao@unr.edu, donzhao@ucdavis.edu
}

\begin{document}

\maketitle

\begin{abstract}
Despite the popularity and practical applicability of blockchains,
there is very limited work on the theoretical foundation of blockchains:
The lack of rigorous theory and analysis behind the curtain of blockchains has severely staggered its broader applications. 
This paper attempts to lay out a theoretical foundation for a specific type of blockchains---the ones requiring basic authenticity from the participants, also called \textit{permissioned blockchain}.
We formulate permissioned blockchain systems and operations into a game-theoretical problem by incorporating constraints implied by the wisdom from distributed computing and Byzantine systems.
We show that in a noncooperative blockchain game (NBG),
a Nash equilibrium can be efficiently found in a closed-form even though the game involves more than two players.
Somewhat surprisingly, the simulation results of the Nash equilibrium implies that the game can reach a stable status regardless of the number of Byzantine nodes and trustworthy players.
We then study a harder problem where players are allowed to form coalitions: the coalitional blockchain game (CBG).
We show that although the Shapley value for a CBG can be expressed in a more succinct form,
its core is empty.

\end{abstract}

\section{Introduction}

Although initially introduced in the form of a cryptocurrency Bitcoin more than a decade ago,
it was not until a few years ago when the internal mechanism named \textit{blockchain} started to draw attention.
While the research and practice on cryptocurrency-driven research on blockchains are surging,
increasingly more proposals are being made on extending blockchains to non-cryptocurrency areas such as smart government, electronic healthcare systems, among many others.
Many systems and frameworks have been developed to mimic and extend the functionality of Bitcoin since then.

Despite the popularity and practical applicability of blockchains,
there is very limited work on the theoretical foundation of blockchains.
While it is a good thing to see blockchain-backed cryptocurrency brings in dynamics to our society, both technically and financially,
it is irresistible to understand the principle based on which blockchains achieves such a success, if there is one such principle.
The lack of rigorous theory and analysis behind the curtain of blockchains has severely staggered its broader applications:
at the end of the day, we need to know not only \textit{whether} and \textit{how} it works, but also \textit{why}.

This paper presents our early attempts to lay out a theoretical foundation for a specific type of blockchains---the ones requiring basic authenticity from the participants, also called \textit{permissioned blockchain}.
We hope the analysis can shed some insights into the principles of blockchains, not only from a computer systems point of view but also human factors.
Admittedly, human factors are somewhat difficult to account in, as many computer system concepts and solutions are hardly applicable.
We thus feel that bringing in the wisdom from utility theory and game theory would be inevitable in order to achieve a deep understanding of blockchains---distributed systems of intelligent agents.
Therefore, we formulate permissioned blockchain systems and operations into a game-theoretical problem by incorporating constraints implied by the wisdom from distributed computing and Byzantine systems.
More specifically, we show that:
\begin{itemize}
    \item In a noncooperative blockchain game (NBG),
a Nash equilibrium can be efficiently found in a closed-form even though the game involves more than two players.
Somewhat surprisingly, the simulation results of the Nash equilibrium implies that the game can reach a stable status regardless of the number of Byzantine nodes and trustworthy players.

    \item We then study a harder problem where players are allowed to form coalitions: the coalitional blockchain game (CBG).
We show that although the Shapley value for a CBG can be expressed in a more succinct form,
its core is empty.
\end{itemize}

In the reminder of this paper,
we will review important concepts and related work in~\S\ref{sec:bg},
articulate the models and assumptions we make for the analysis in~\S\ref{sec:model},
define and analyze noncooperative and coalitional blockchains in~\S\ref{sec:nbg} and~\S\ref{sec:cbg}, respectively,
and finally conclude the paper in~\S\ref{sec:conclusion},

\section{Background and Related Work}
\label{sec:bg}

\subsection{Blockchains}

Although somewhat being mystified especially through the hype of cryptocurrency like Bitcoin,
the technical idea of blockchain is surprisingly simple.
From a data structure standpoint, a blockchain is essentially a set of replicated \textit{linkedlist}s of blocks:
each block is verified by the descendent block through a hash function. 
From a system perspective, each copy of these linkedlists is usually deployed on a distinct machine, being a physical computer or virtual machine;
the tricky part is, however, how to \textit{synchronize} these linkedlists,
and that is exactly the root cause of many challenges, such as consistency, performance, and security.
Specifically, there are a series of questions we need to answer:
\textit{When} should these lists be synchronized?
\textit{How} can we verify the lists have been synchronized?
\textit{Where} would these synchronizations happen?
just to name a few.
\textit{Synchronization} is one of the oldest problems in many system perspectives, and thus recently renewed by the popularity of blockchains.
Various communities showed a lot of interests in blockchains, notably in distributed systems~\cite{mherlihy_cacm19}, database systems~\cite{pruan_vldb19}, operating systems~\cite{jlind_sosp19}, and network systems~\cite{Wang_nsdi19}.

Synchronization also partly impacts the categorization of blockchains. 
While the popular view of categorizing blockchains is to whether open the blockchain to the public, leading to the concept of \textit{permissionless blockchains} and \textit{permissioned blockchains},
we want to point out that an orthogonal, if not more understandable, criterion to classify a blockchain system lies on its underlying synchronization mechanism.
At one extreme, the synchronization can be completely based on individualism, being represented by Proof-of-Work (PoW)~\cite{bitcoin}, Proof-of-Stake (PoS)~\cite{ethereum}, among many others.
The commonality of these mechanisms is that a single entity in the blockchain, somehow, declares its superiority over others in a competing round, and the whole blockchain moves forward under its leadership.
At the other end of the spectrum, we observe a more democratic implementation of synchronization: every node in the blockchain network has a vote about how to proceed to the next step.
At the writing of this paper, the \textit{de facto} protocol for ensuring this collaborative synchronization is practical Byzantine fault tolerance (PBFT)~\cite{pbft_osdi99} addressing the original problem~\cite{llamport_tpls82}.

\subsection{Byzantine Distributed Systems}

Although widely used in literature, the term \textit{distributed system} is seldom formally defined. 
Here we adopt the definition from~\cite{steen_book13}:
``\textit{A distributed system is a collection of autonomous computing elements that appears to its users as a single coherent system}.''
One of the most challenging problems in distributed systems is fault tolerance:
how would the distributed system mask its internal failures such that the user cannot notice the failures and, more importantly, the dispatched tasks are correctly and efficiently completed?
To answer this question, we need first to define what a \textit{failure} really refers to.
While there are various ways to differentiate types of failures,
the most notable types are \textit{crash failures} and \textit{Byzantine failures}:
the former refers to those failures that cause node crash, and the latter refers to arbitrary failures including but not limited to malicious attacks that allow the node to continue to operate with (intentionally) wrong data and states.  
The name ``Byzantine'' originates from warlords' treacherous behaviors in the ancient Byzantine Empire.
In this paper, we use \textit{Byzantine Distributed Systems} to refer to those distributed systems which adopt strong consensus protocols to tolerate Byzantine failures.

Byzantine distributed systems are considered one of the hardest paradigms,
as the behavior of those Byzantine nodes is arbitrary and cannot be predicted or estimated.
This is even harder than the well-known fail-stop paradigm, 
which can be effectively circumvented by Paxos~\cite{llamport_tcs98} and its variants. 
It was not until 1999 when the first practical protocol~\cite{pbft_osdi99} for tolerating Byzantine failures was published.
In~\cite{pbft_osdi99}, the authors demonstrated that it was not only possible but also feasible to have non-faulty nodes more than twice as the number of Byzantine nodes such that the overall system was consistent following the proposed 3-phase protocol called practical Byzantine fault tolerance (PBFT).
Since then, many follow-up improvements such as reducing the number of nonfaulty nodes and reducing the number of phases have been proposed;
and yet, the original PBFT protocol is still considered as a good balance between generality and simplicity, and thus widely used in practice (e.g., Hyperledger Fabric~\cite{hyperledger_eurosys18}.

\subsection{Game Theory}

Game theory was established to mathematically model many practical problems in the form of games.
It was first applied to macroeconomics to explain many individual behaviors and then quickly extended to other disciplines such as politics, social science, biological science, and computer science.
For example, auction theory---a subarea in game theory---has been widely used in radio spectrum allocation.
One of the key insights brought by game theory is the introduction to rationality:
each participant, or more commonly called an \textit{agent} in the game theory literature, is \textit{intelligent}, meaning that she has her own interest, such as maximizing a utility.
This is in sheer contrast to distributed systems, which simply assumes the nodes are \textit{machines}.\footnote{Indeed, one can argue that a machine can also be \textit{intelligent}.}

If we rethink blockchains, each node is, in fact, a mix of a machine \textit{and} an agent.
Put it in another way, a blockchain node can be Byzantine, or can be maximizing the user's utility, or both.
There was a good survey work~\cite{iabraham_sigact11} on discussing the commonality and difference between game theory and distributed computing.
More recently, applying game-theoretical approaches~\cite{ittay_ccs18,ittay_sp15} to blockchain security also emerged.
This paper, to our knowledge, is the first systematic work on blockchains from both the distributed computing and the game-theoretical perspectives.

\section{Models and Assumptions}
\label{sec:model}


There are three type of nodes, or players, involved in a blockchain system:
a good citizen (\texttt{C}) who always votes for the proposal,
a terrorist (\texttt{T}) who always votes against the proposal,
and an adventurer (\texttt{A}) who makes her decision to maximize the utility (either voting for or against the proposed value, i.e., $\mathtt{A} = \mathtt{A_g} \cup \mathtt{A_b}$).
We use $N = \mathtt{C} \cup \mathtt{T} \cup \mathtt{A}$ to denote the entire set of nodes, and use $\mathbf{n} = |N|$ to indicate the cardinality of the node set.

\textbf{Terrorist.} 
The term \textit{terrorist} first appeared in the pioneer work~\cite{tmosc_podc06},
which, for the first time, synthesized two orthogonal research communities of distributed computing and game theory in the famous \textit{virus inoculation game}.
Specifically, a \textit{terrorist} is irrational in the sense that he does not follow any protocol and is disinterested in either the utility or the cost (for the attack).
From a distributed computing standpoint, a terrorist is definitely a \textit{faulty} node that will vote against the proposed (true) value.
Conventionally, such nodes are called Byzantine nodes whose actions are completely arbitrary, including coalition with other peers against the proposed value.
In blockchains, a terrorist could be, for example, a malicious hacker intentionally forking a long sidechain, 
a proxy node from another blockchain competitor to slow down the performance, to name a few.
We will use $\mathbf{t} = |\mathtt{T}|$ to denote the cardinality of the terrorist set $\mathtt{T}$.

\textbf{Good Citizen.}
In contrast to a terrorist, a good citizen (or, a citizen for short in the following discussion) \textit{always} votes for the proposed value.
From a distributed system's standpoint, a citizen represents a highly reliably node with automatic backup and recovery,
meaning that it follows the consensus protocol for the eventual agreement across participating nodes.
From a game theory's point of view, however,
a citizen is not considered rational in the sense that she does not attempt to maximize any utility and only follow the consensus protocol---she is \textit{not} self-interested.
We will use $\mathbf{c} = |\mathtt{C}|$ to denote the cardinality of the citizen set $\mathtt{C}$.

\textbf{Adventurer.}
An adventurer is an opportunist:
she will maximize her utility by voting for or against the proposed value.
Nevertheless, her utility is dependent not only on her own action but also others'---a \textit{normal-form game} because:

\begin{itemize}
    \item \texttt{(Finite Players)} All nodes in a blockchain are identifiable using their public keys; 
    
    \item \texttt{(Finite Actions)} Each node has three available actions on the new proposed value: \textit{yes}, \textit{no}, and \textit{abstain}. 
    Each of the actions leads to a numerical payoff.
\end{itemize}

Depending on the votes made by an adventurer, we further denote the set of adventurers voting \texttt{Yes} as $\mathtt{A_g}$ and the set of the other adventurers who vote \texttt{No} as $\mathtt{A_b}$.
In the following discussion, we use
$\mathbf{g} = |\mathtt{A_g|}$ and $\mathbf{b} = |\mathtt{A_b|}$
for the two subsets' cardinalities, respectively.

\section{Noncooperative Blockchain Game}
\label{sec:nbg}

\subsection{Definition}

\subsubsection{Strategy Profiles}

A distributed system including blockchains usually adopts a \textit{timeout} mechanism that set some unresponsive nodes to a default value, such as \textit{null} or \textit{nil}, as an ``abstention'' vote.
From a pure system point of view, an abstention vote is not different than a \textit{no} vote---the system usually takes a very conservative position in interpreting the responses.
In the case of permissioned blockchains, therefore, the consensus protocol interprets an \textit{abstention} vote as a \textit{no} vote.
Formally, we have two pure strategies (or, actions) for all the $\mathbf{n}$ nodes in the blockchain: 
$\{\mathtt{Yes}, \mathtt{No}\}$, $0 \le i < \mathbf{n}$.
It should be noted that a strategy in game theory can be mixed:
although we have only two \textit{pure} strategies here,
the set of statistical distributions, denoted by $S_i$ for the $i$-th player, over only two values is still considerable and, arguably, more importantly, serves as a cornerstone for the Nash equilibrium:
the high-dimensional simplex theory and fixed-point theorems are all built upon the idea of convex combinations, which, implicitly, assumes a ``mix'' of strategies.

If the system is not compromised, the non-faulty nodes will continue to work on ``agreeing'' on the next proposed value and get rewarded by a transaction fee, and the faulty nodes might be forced to leave the network;
Otherwise, the faulty nodes overturn the existing network and collect a high reward, leaving the ``honest'' nodes' work worthless.
Note that we ignore the cost (e.g., electricity, hardware procurement, space rental) for mining blocks or attacking the network.
Formally, the utility function $u_i$ for a player $i$ is defined as follows:
\begin{equation}
\label{eq:payoff1}
    u_n = 
\begin{cases}
    \frac{\displaystyle p_n}{\displaystyle \mathbf{g} + \mathbf{c}},            & \text{if } \mathbf{g} + \mathbf{c} > \mathbf{b} + \mathbf{t}\\
    0,              & \text{otherwise}
\end{cases}
\end{equation}
where $p_n$ denotes the overall payoff of a winning ``honest'' nodes and $n$ denotes that winning non-faulty node, i.e., $n \in \mathtt{A_g} \cup \mathtt{C}$;
and
\begin{equation}
\label{eq:payoff2}
    u_f= 
\begin{cases}
    \frac{\displaystyle p_f}{\displaystyle \mathbf{b}}, 
    & \text{if } \mathbf{g} + \mathbf{c} \le \mathbf{b} + \mathbf{t}\\
    0,              & \text{otherwise}
\end{cases}
\end{equation}
where $p_f$ denotes the overall payoff of all ``deviating'' nodes and $f$ denotes a faulty node, i.e., $f \in \mathtt{A_b}$.

In practice, $p_f$ is orders of magnitude higher than $p_n$,
motivating those adversaries to take the risk.
As a concrete example, an honest node in a Bitcoin network is rewarded 12.5 BTCs,
which, at the writing of this paper, are roughly worth \$93,000 USD\footnote{We do not count the transaction fee, usually about \$0.1x USD per transaction.};
however, if the network is compromised, the net loss could be in the order of tens of millions of dollars.
For example, 72-million dollars were stolen in the notorious mega-hack in 2016~\cite{bitcoin_megahack}.

\subsubsection{Zero-Sum among Non-Byzantine Nodes}

An important observation is that all the non-Byzantine nodes constitute a zero-sum game.
That is, the Byzantine nodes, or terrorists, are disinterested in the expense or utility incurred in the game---the only objective is to compromise the network.
As a consequence, we have the following zero-sum constraint:
\begin{equation}
\label{eq:zerosum}
    0 \equiv \mathbf{c} \cdot p_n + \mathbf{g} \cdot p_n - \mathbf{b} \cdot p_f
\end{equation}
where the first term indicates the payoff for good citizen,
the second term indicates the payoff for adventurers who vote for the proposal,
the third term indicates the payoff for adventurers who vote against the proposal.

\subsubsection{Consensus Protocols}
\label{subsec:game_consensus}

There are rich literature in handling \textit{arbitrary} nodes in the distributed system community. 
In the context of permissioned blockchains (Hyperledger Fabric~\cite{hyperledger_eurosys18} and other variants based on PBFT~\cite{pbft_osdi99}),
the state-of-the-art takes a quorum-based mechanism to move forward.
In permissioned blockchains, 
the consensus protocols require that non-faculty nodes outnumber the faulty nodes by at least 200\%:
$\mathbf{g} + \mathbf{c} > 2(\mathbf{b} + \mathbf{t})$, or more commonly known as
$\mathbf{n} \geq 3\cdot (\mathbf{b}+\mathbf{t}) + 1$.
More formally, we set the difference as one between the two groups as the borderline case in the following discussion:
\begin{equation}
\label{eq:consensus_private}
    \mathbf{c} + \mathbf{g} \geq 1 + 2 \cdot (\mathbf{t} + \mathbf{b})
\end{equation}

\begin{definition}[Noncoorperative blockchain game]
A noncooperative Byzantine game represented by a tuple $\langle N, S, u, P \rangle$,
where $N$ represents the players, $S = S_1 \times \cdots \times S_\mathbf{n}$ denotes the mixed-strategy profiles, $u$ indicates the utility functions,
and $P$ denotes the consensus protocol among the players,
such that each adventurer node maximizes its utility without violating $P$.
\end{definition}

Essentially, the \textit{free} players would choose between being part of either $\mathtt{A_g}$ or $\mathtt{A_b}$, exclusively,
such that her utility is maximized under the constraints specified by Equation~\ref{eq:payoff1},
Equation~\ref{eq:payoff2},
Equation~\ref{eq:zerosum},
and Equation~\ref{eq:consensus_private}.

\begin{figure}[!t]

  \centering
  \includegraphics[width=85mm]{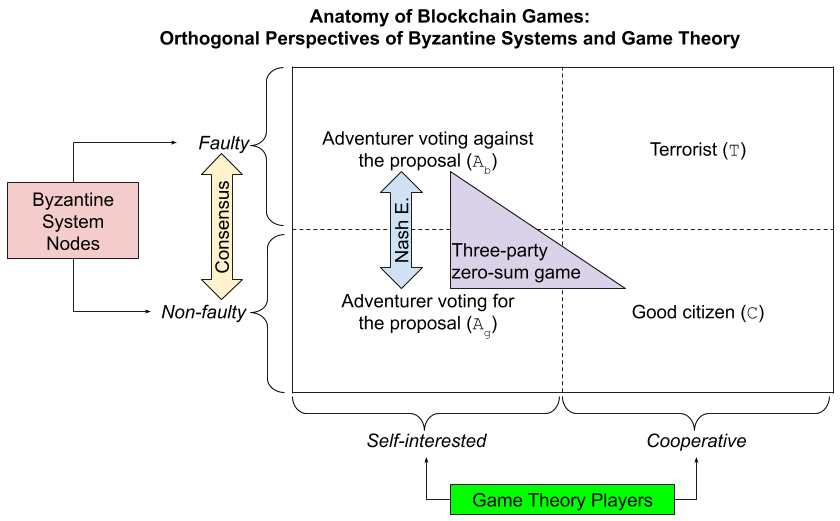}
  \caption{
    Noncooperative blockchain game.
}
    \label{fig:diagram}  
\end{figure}

We illustrate the whole game in Figure~\ref{fig:diagram}.
We conjecture that a noncooperative blockchain game (NBG) exists a Nash equilibrium between $\mathtt{A_g}$ and $\mathtt{A_b}$,
in which no player from one set has the incentive to change her mind to move to the other set for higher payoff---the cardinalities of $\mathtt{A_g}$ and $\mathtt{A_b}$ reach equilibrium under unilateral change of strategies.
The next section will test this hypothesis and,
in fact, demonstrate that there is indeed a Nash equilibrium in a closed-form.
Other solution concepts are also possible, and yet beyond the scope of this paper.

\subsection{Nash Equilibrium}

Since the goal is to analyze the decisions made by $\mathtt{A_g}$ and $\mathtt{A_b}$,
in the following discussion, we assume the other variables are known.
That is, we assume the values of $\mathbf{c}$ and $\mathbf{t}$ are well estimated before the game starts.
Also, we assume the payoff values $p_n$ and $p_f$ can be accurately predicted as \textit{a priori}.

If we consider the system of Equation~\ref{eq:zerosum} and Equation~\ref{eq:consensus_private}, i.e.,
\begin{equation}
\label{eq:equil_system}    
\begin{cases} 
\mathbf{c} \cdot p_n + \mathbf{g} \cdot p_n - \mathbf{b} \cdot p_f = 0 \\ 
\mathbf{c} + \mathbf{g} \geq 1 + 2 \cdot (\mathbf{t} + \mathbf{b}) 
\end{cases}
\end{equation}
we will have
\[
\mathbf{b = (\mathbf{g} + \mathbf{c})\cdot \gamma}
\]
where $\gamma = \frac{\displaystyle p_n}{\displaystyle p_f}$.
Note that in the real world, $p_f$ is orders of magnitude larger than $p_n$, implying that $0 < \gamma \ll 1$.
We call this variable \textit{reciprocal risk factor} (RRF),
indicating the payoff ratio of a compliant action over a deviating action.

We can then rewrite the inequality in Eq.~\ref{eq:equil_system} into
\begin{align*}
\mathbf{g} &\ge 1 + 2(\mathbf{t} + \gamma\cdot(\mathbf{g} + \mathbf{c})) - \mathbf{c}\\
&= 1 + 2\mathbf{t} + (2\gamma-1)\cdot \mathbf{c} + 2\gamma\cdot \mathbf{g}
\end{align*}
Consequently, we derive $\mathbf{g}$ in the following form:
\begin{equation}
\label{eq:equil_good}
    \mathbf{g} \geq \frac{1 + 2\mathbf{t}}{1 - 2\gamma} - \mathbf{c}
\end{equation}
That is, to satisfy all the game constraints, 
the number of \textit{good} adventurers has to exceed a specific threshold jointly determined by the number of good citizens, the number of terrorists, and RRF.

Similarly, we can derive $\mathbf{b}$ in the following form:
\begin{equation}
\label{eq:equil_bad}
    \mathbf{b} \geq \frac{\gamma\cdot(1 + 2\mathbf{t})}{1 - 2\gamma}
\end{equation}
Interestingly, the number of \textit{bad} adventurers is independent of the number of citizens,
even though there is a zero-sum constraints involving both groups.

In practice, the overall payoff of the winning side,
either the faulty cohort or the non-faulty cohort,
is fixed.
Therefore, both $\mathbf{g}$ and $\mathbf{b}$ would be preferably set to the minimums.
As a result, we have the following optimal setup:
\begin{equation}
\label{eq:equil_result}    
\begin{cases} 
\mathbf{g^*} = \frac{\displaystyle 1 + 2\mathbf{t}}{\displaystyle 1 - 2\gamma} - \mathbf{c} \\
\mathbf{b^*} = \frac{\displaystyle \gamma\cdot(1 + 2\mathbf{t})}{\displaystyle 1 - 2\gamma}
\end{cases}
\end{equation}
That is, if there are $\mathbf{g}^*$ \textit{good} adventurers and $\mathbf{b}^*$ \textit{bad} adventurers in the game,
then the all constraints are satisfied.
We then calculate the ratio of $\mathbf{g}^*$ over $\mathbf{g}^* + \mathbf{b}^*$:
\begin{equation}
\label{eq:nash_equil} 
Pr(g) = \frac{\displaystyle 1 + 2\mathbf{t} - (1 + 2\gamma)\mathbf{c}}{\displaystyle 1 + \gamma + 2(1 + 2\gamma)(2\mathbf{t} - \mathbf{c})}
\end{equation}
Let $Pr(g)$ be the probability of an adventurer voting for \texttt{Yes}.
That is, the mixed strategy $s^*$ of an adventurer expressed in a lottery is $l = [Pr(g): \text{Yes}, 1-Pr(g): \text{ No}]$.

\begin{theorem}
\label{thm:nash_equil}
Given a noncoorperative blockchain game,
a mixed strategy with probability $Pr(g)$ of voting \texttt{Yes} is a Nash equilibrium.
\end{theorem}

\begin{proof}
The scratch of our proof is as follows.
It should be noted that the summation of $\mathbf{g}$ and $\mathbf{b}$ is constant.
Therefore, an adventurer unilaterally changing her local strategy would decrement its the cardinality of her original set (either $\mathtt{A_g}$ or $\mathtt{A_b}$) and simultaneously increment the other one by one.
It then boils down to two cases:
(i) a ``bad'' adventurer changes to a ``good'' one, or
(ii) a ``good'' adventurer changes to a ``bad'' one.
We will investigate both cases in the following.
For the sake of brevity, we will refer to Eq.~\ref{eq:equil_result} instead of Eq.~\ref{eq:nash_equil} in the proof.

The first case is trivial. An adventurer $a$ in $\mathtt{A_b}$ would not bother to change his decision to join $\mathtt{A_g}$ because the new payoff would be lower anyhow,
i.e., $p_n < p_f$.
That is, assuming all of the other players in $\mathtt{A} \textbackslash \{a\}$ do not change their votes,
$a$ would not increase her utility unilaterally.

The second case, however, needs a bit more work because in contrast to the first case,
now an adventurer $a'$ has a good reason to move from $\mathtt{A_g}$ to $\mathtt{A_b}$ in pursue of higher utility under the condition that all other players do not change their strategies and the system still functions as before under the constraints including zero-sum and consensus protocols.
We assume now $a'$ does move from $\mathtt{A_g}$ to $\mathtt{A_b}$ from previous state $E$;
then the new numbers of ``good'' ($\hat{\mathbf{g}}$) and ``bad'' ($\hat{\mathbf{b}}$) adventurers, are
\[
\begin{cases} 
\hat{\mathbf{g}} = \mathbf{g^*} - 1 = 2\cdot \frac{\displaystyle \gamma + \mathbf{t}}{\displaystyle 1 - 2\gamma} - \mathbf{c} 
\\
\hat{\mathbf{b}} = \mathbf{b^*} + 1 = \frac{\displaystyle 1 + \gamma\cdot(2\mathbf{t} - 1)}{\displaystyle 1 - 2\gamma}
\end{cases}
\]
And we know that $\hat{E} = \langle \hat{\mathbf{g}}, \hat{\mathbf{b}} \rangle$ will break the consensus protocol requirement specified in Eq.~\ref{eq:consensus_private}.
In this case, from a distributed system point of view,
there will be another player $a''$ attempting to move from $\mathbf{A_b}$ to $\mathbf{A_g}$ because otherwise $a''$ will receive zero payoff if the system reboots\footnote{So as other players in $\mathbf{A_b}$, although $a''$ does not care.} due to the Byzantine failure;
$p_n$ is still better than nothing.
Therefore, $\mathbf{g}$ and $\mathbf{b}$ will be reset to $\mathbf{b^*}$ and $\mathbf{b^*}$, respectively.
\end{proof}

\subsection{Simulation Results}

We apply the model and equilibrium on a sample 10,000-node blockchain.
Note that the Bitcoin network currently comprises about 10,000 miners worldwide~\cite{bitcoin}.
We set $p_n$ = 100,000 and $p_f$ = 70,000,000.
Again, these are practical numbers in the real world.~\cite{bitcoin,bitcoin_megahack}.

\begin{figure*}[!t]
    \centering
    \includegraphics[width=150mm]{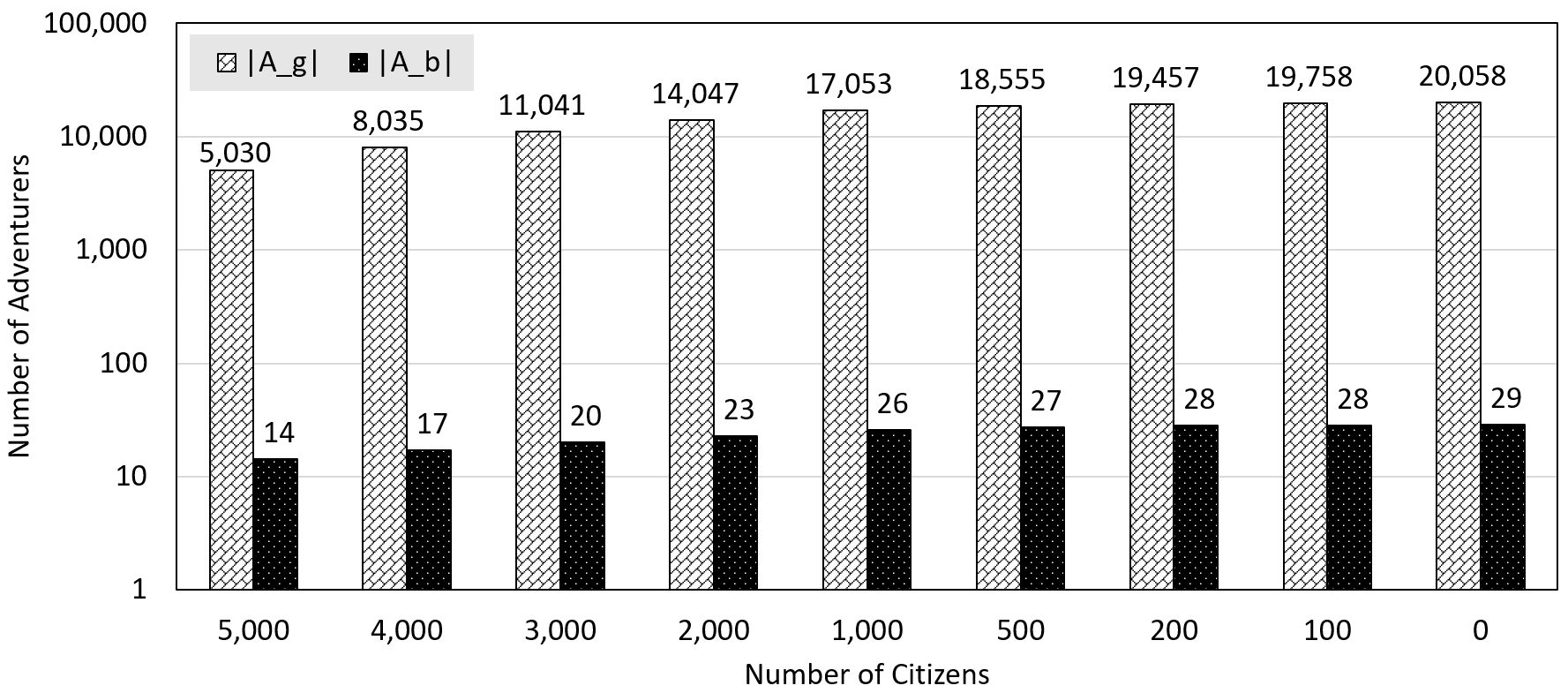}
    \caption{Simulation result on a 10,000-node blockchain.}
    \label{fig:result}
\end{figure*}

Figure~\ref{fig:result} reports the number of adventurers corresponding to various portions of good citizens.
Interestingly, the number of adversarial adventurers stays at a stable level regardless of the dramatic change of portions between citizens and terrorists.
On the other hand, good-will adventurers will quickly fill out the space left by the missing citizens.
The implication of this result is intriguing:
as long as we can open the market to a large pool of players (i.e., adventurers),
the system will reach equilibrium even if there are zero good citizens in the game.

\section{Coalitional Blockchain Game}
\label{sec:cbg}

\subsection{Definition}

This section sheers our focus from noncooperative normal-form games into the more complex cooperative games, also known as coalitional games,
which essentially allows the coalition form by subsets of players.
In other words, there will be a grand coalition formed by all the non-Byzantine players: $\mathtt{A_g}$, $\mathtt{A_b}$, and $\mathtt{C}$.
Essentially, when we assume individual players can collaborate in a coalitional game,
the groups formed by the individual players become the granularity in the analysis.
Consequently, when there is no confusion in the context, the term \textit{player} in a coalitional game really refers to a group of individual players in the sense of noncooperative games.

As all the coalitional games, we will strive to answer the following two questions:
(i) What is the Shapley value for each player to feel the game is fair?
(ii) Is the core empty? If not, can we efficiently find one such that all players reach equilibrium?
Before answering both questions, we first define the \textit{coalitional blockchain game} (CBG) as follows.

\begin{definition}[Coalitional blockchain game]
\label{def:cbg_game}
A coalitional blockchain game is a pair $\langle N, v \rangle$, where $N$ is a set of three players $\mathbf{n} = 3$ and a characteristic function $\displaystyle v: 2^N \mapsto \mathbb{R}$ associates with each subset of $N$ with a payoff, denoted by $v(S)$ for all $S \subseteq N$:
\begin{equation}
\label{eq:payoff_coalition}
    v(S) = 
\begin{cases}
    p,            & \text{if } |S| > \frac{n}{2} \\
    0,              & \text{otherwise}
\end{cases}
\end{equation}
where $p > 0$ indicates the total payoff of the game.
\end{definition}

\subsection{Shapley Value}

The Shapley value characterizes the fairness perceived by each player:
it quantifies the division of the grand-coalition payoff.
Formally, the general form of the Shapley value for the $i$-th player calculated by
\[
\phi_i = \frac{1}{\mathbf{n!}} \cdot \sum_{S \subseteq N \textbackslash \{i\}} \overbrace{\mathbf{s}! \cdot (\mathbf{n} - \mathbf{s} - 1)!}^{\text{combinatorial coefficient}} \cdot \underbrace{(v(S \cup \{i\}) - v(S))}_{\text{marginal contribution}}
\]
where $\mathbf{s} = |S|$.
A singleton coalitional blockchain game has only three players;
therefore, the factorials can be efficiently calculated,
which we will do soon.
However, we want to point it out that even for a ``meta'' version of coalition blockchain games---a player (or, a group) is involved in multiple, independent blockchain games,
the factorials can still be accurately estimated by the Stirling approximation.
The term $v(S \cup \{i\}) - v(S)$ is also called \textit{marginal contribution} and therefore, the Shapley can be intuitively understood as the \textit{combinatorial average} of each player's marginal contribution.
In the following, we will dive deeper into this concept in the context of coalitional blockchain games and derive a more succinct and meaningful expression of the Shapley value.

According to Equation~\ref{eq:payoff_coalition}, the marginal contribution of each player can be only $p$ or 0:
it degenerates to a binary variable,
or more commonly known as a \textit{dummy variable} denoted by $D_i(S) \in \{1,0\}$:
\begin{equation}
\label{eq:dummy}   
    D_i(S) = 
\begin{cases}
    1,  & \text{if } |S| + 1 > \frac{n}{2} \text{ and } |S| \le \frac{n}{2} \\
    0,  & \text{otherwise}
\end{cases}
\end{equation}
Therefore, the marginal contribution can be rewritten as $p \cdot D_i(S)$.

In the defined coalitional blockchain game, we always have $\mathbf{n} = 3$.
Therefore, we have $0 \le \mathbf{s} \le 2$, and $0 \le \mathbf{n} - \mathbf{s} - 1 \le 2$.
We can construct the following (proposigional logic) formula:
\begin{equation}
\label{eq:cc}    
E(S) = \neg (|S| \text{ mod } 2) + 1
\end{equation}
It is easy to verify the equivalence between $E(S)$ and the \textit{combinatorial coefficient} in the general definition of Shapley value.
Therefore, the Shapley value of a coalitional blockchain game is defined as:
\begin{equation}
\label{eq:shapley_cbg} 
\phi^{\text{CBG}}_i \triangleq {\displaystyle \frac{p}{6}} \cdot \sum_{S \subseteq N \textbackslash \{i\}} E(S) \cdot D_i(S)
\end{equation}

\subsection{Core}

Unlike the Nash equilibrium that always exists in a noncooperative game,
the counterpart equilibrium, namely the core, in a coalitional game, may not exist.
Even worse, the problem of deciding whether an arbitrary coalitional game has an empty core is, in many cases, NP-complete.
In the remainder of this section, we will show that a coalitional blockchain game has an empty core.

First, we reiterate important definitions in coalitional game theory.
A \textit{core} in a coalitional game is a set of payoff vectors in the grand coalition such that no smaller coalition collectively can do better:
\[
\forall S \subseteq N, \sum_{i \in S} x_i \geq v(S)
\]
where $x_i$ is the payoff function for player $i$.
To show the emptiness of the core, we also need to define a class of coalitional game called \textit{additive coalitional game} as follows.
\begin{definition}[Additive coalitional game]
\label{def:additive_game}
A coalitional game $\langle N, v \rangle$ is \textit{additive} if for any two distinct subsets $S \subseteq N$, $T \subseteq N$, and $S \cap U = \emptyset$, such that $v (S \cup U) = v(S) + v(U)$.
\end{definition}

\newtheorem{lemma}[theorem]{Lemma}
\begin{lemma}
\label{lemma:cbg_nonadditive}
A coalitional blockchain game is not additive.
\end{lemma}

\begin{proof}
It is sufficient to prove the lemma by identifying one counterexample.
Let $\mathbf{g} = 1$ and $\mathbf{c} = \frac{n}{2}$.
We then have $v(\mathtt{A_g}) = 0$ and $v(\mathtt{C}) = 0$.
By definition, however, $v(\mathtt{A_g} \cup \mathtt{C}) = p > 0 \not= v(\mathtt{A_g}) + v(\mathtt{C})$.
\end{proof}

\begin{theorem}
\label{thm:cbg_emptycore}
A coalitional blockchain game has an empty core.
\end{theorem}

\begin{proof}
First, we note that the coalitional blockchain game is a \textit{constant-sum} game,
because the overall balance among the three players does not change after the game.
On the other hand, according to Lemma~\ref{lemma:cbg_nonadditive}, we cannot simply calculate the overall payoff of a coalition of multiple players---that is, there exists interference among players.
As a result, the grand payoff would not be \textit{always} larger than or equal to the balance-weighted-payoffs from subsets of the grand coalition.
According to the Bondereva-Shapley theorem, such a coalitional game does not have a non-empty core.
Alternatively, the same conclusion can be drawn by applying the well-known result that states that ``a constant-sum coalitional game has no nonempty core if the game is not additive.''~\cite{yshoham_mas08}
\end{proof}

\section{Conclusion and Future Work}
\label{sec:conclusion}

This paper attempts to lay out a theoretical foundation for a specific type of blockchains---the ones requiring basic authenticity from the participants, also called \textit{permissioned blockchain}.
We formulate permissioned blockchain systems and operations into a game-theoretical problem by incorporating constraints implied by the wisdom from distributed computing and Byzantine systems.
We show that in a noncooperative blockchain game (NBG),
a Nash equilibrium can be efficiently found in a closed-form even though the game involves more than two players.
Somewhat surprisingly, the simulation results of the Nash equilibrium implies that the game can reach a stable status regardless of the number of Byzantine nodes and trustworthy players.
We then study a harder problem where players are allowed to form coalitions: the coalitional blockchain game (CBG).
We show that although the Shapley value for a CBG can be expressed in a more succinct form,
its core is empty.
Our main future work is to extend the game into an extensive form:
the blockchain will be modeled as a series of rounds.

\newpage
\bibliographystyle{named}
\bibliography{ijcai20}

\begin{thebibliography}{}

\bibitem[\protect\citeauthoryear{Abraham \bgroup \em et al.\egroup
  }{2011}]{iabraham_sigact11}
Ittai Abraham, Lorenzo Alvisi, and Joseph~Y. Halpern.
\newblock Distributed computing meets game theory: Combining insights from two
  fields.
\newblock {\em SIGACT News}, 2011.

\bibitem[\protect\citeauthoryear{Androulaki \bgroup \em et al.\egroup
  }{2018}]{hyperledger_eurosys18}
Elli Androulaki, Artem Barger, Vita Bortnikov, Christian Cachin, Konstantinos
  Christidis, Angelo De~Caro, David Enyeart, Christopher Ferris, Gennady
  Laventman, Yacov Manevich, Srinivasan Muralidharan, Chet Murthy, Binh Nguyen,
  Manish Sethi, Gari Singh, Keith Smith, Alessandro Sorniotti, Chrysoula
  Stathakopoulou, Marko Vukoli\'{c}, Sharon~Weed Cocco, and Jason Yellick.
\newblock Hyperledger fabric: A distributed operating system for permissioned
  blockchains.
\newblock In {\em Proceedings of the Thirteenth EuroSys Conference}, EuroSys
  '18, pages 30:1--30:15, New York, NY, USA, 2018. ACM.

\bibitem[\protect\citeauthoryear{{Bitcoin Mega-Hack}}{2016}]{bitcoin_megahack}
{Bitcoin Mega-Hack}.
\newblock \url{
  https://proeconomia.wordpress.com/2016/08/05/weekly-globo-digest}, 2016.

\bibitem[\protect\citeauthoryear{{Bitcoin}}{2008}]{bitcoin}
{Bitcoin}.
\newblock \url{ https://bitcoin.org/bitcoin.pdf}, 2008.

\bibitem[\protect\citeauthoryear{Castro and Liskov}{1999}]{pbft_osdi99}
Miguel Castro and Barbara Liskov.
\newblock Practical byzantine fault tolerance.
\newblock In {\em Proceedings of the Third Symposium on Operating Systems
  Design and Implementation}, OSDI '99, pages 173--186, Berkeley, CA, USA,
  1999. USENIX Association.

\bibitem[\protect\citeauthoryear{{Ethereum}}{2019}]{ethereum}
{Ethereum}.
\newblock \url{ https://www.ethereum.org/}, 2019.

\bibitem[\protect\citeauthoryear{Eyal}{2015}]{ittay_sp15}
Ittay Eyal.
\newblock The miner's dilemma.
\newblock In {\em IEEE Symposium on Security and Privacy (S\&P)}, pages
  89--103, 2015.

\bibitem[\protect\citeauthoryear{Herlihy}{2019}]{mherlihy_cacm19}
Maurice Herlihy.
\newblock Blockchains from a distributed computing perspective.
\newblock {\em Commun. {ACM}}, 62(2):78--85, 2019.

\bibitem[\protect\citeauthoryear{Lamport \bgroup \em et al.\egroup
  }{1982}]{llamport_tpls82}
Leslie Lamport, Robert Shostak, and Marshall Pease.
\newblock The byzantine generals problem.
\newblock {\em ACM Trans. Program. Lang. Syst.}, 4(3):382–401, July 1982.

\bibitem[\protect\citeauthoryear{Lamport}{1998}]{llamport_tcs98}
Leslie Lamport.
\newblock The part-time parliament.
\newblock {\em ACM Trans. Comput. Syst.}, 16(2):133–169, May 1998.

\bibitem[\protect\citeauthoryear{Lind \bgroup \em et al.\egroup
  }{2019}]{jlind_sosp19}
Joshua Lind, Oded Naor, Ittay Eyal, Florian Kelbert, Emin~G\"{u}n Sirer, and
  Peter Pietzuch.
\newblock Teechain: A secure payment network with asynchronous blockchain
  access.
\newblock In {\em Proceedings of the 27th ACM Symposium on Operating Systems
  Principles (SOSP)}, page 63–79, 2019.

\bibitem[\protect\citeauthoryear{Moscibroda \bgroup \em et al.\egroup
  }{2006}]{tmosc_podc06}
Thomas Moscibroda, Stefan Schmid, and Rogert Wattenhofer.
\newblock When selfish meets evil: Byzantine players in a virus inoculation
  game.
\newblock In {\em Proceedings of the Twenty-fifth Annual ACM Symposium on
  Principles of Distributed Computing}, PODC '06, pages 35--44, New York, NY,
  USA, 2006. ACM.

\bibitem[\protect\citeauthoryear{Ruan \bgroup \em et al.\egroup
  }{2019}]{pruan_vldb19}
Pingcheng Ruan, Gang Chen, Anh Dinh, Qian Lin, Beng~Chin Ooi, and Meihui Zhang.
\newblock Fine-grained, secure and efficient data provenance for blockchain.
\newblock {\em {PVLDB}}, 12(9):975--988, 2019.

\bibitem[\protect\citeauthoryear{Shoham and Leyton-Brown}{2008}]{yshoham_mas08}
Yoav Shoham and Kevin Leyton-Brown.
\newblock {\em Multiagent Systems: Algorithmic, Game-Theoretic, and Logical
  Foundations}.
\newblock Cambridge University Press, USA, 2008.

\bibitem[\protect\citeauthoryear{Tanenbaum and Van~Steen}{2013}]{steen_book13}
Andrew~S. Tanenbaum and Maarten Van~Steen.
\newblock {\em Distributed Systems}.
\newblock Pearson Education, 2013.

\bibitem[\protect\citeauthoryear{Tsabary and Eyal}{2018}]{ittay_ccs18}
Itay Tsabary and Ittay Eyal.
\newblock The gap game.
\newblock In {\em Proceedings of the 2018 ACM SIGSAC Conference on Computer and
  Communications Security (CCS)}, page 713–728, 2018.

\bibitem[\protect\citeauthoryear{Wang and Wang}{2019}]{Wang_nsdi19}
Jiaping Wang and Hao Wang.
\newblock Monoxide: Scale out blockchain with asynchronized consensus zones.
\newblock In {\em 16th {USENIX} Symposium on Networked Systems Design and
  Implementation (NSDI)}, 2019.

\end{thebibliography}

\end{document}